\documentclass[fleqn,usenatbib]{mnras}

\usepackage{newtxtext,newtxmath}

\usepackage[T1]{fontenc}

\DeclareRobustCommand{\VAN}[3]{#2}
\let\VANthebibliography\thebibliography
\def\thebibliography{\DeclareRobustCommand{\VAN}[3]{##3}\VANthebibliography}

\usepackage{graphicx}	
\usepackage{amsmath}	

\title[flavour conversion in the DNNB]{Impact of Neutrino Flavour Conversion on the Diffuse Neutrino Background from Neutrino-dominated Accretion Flows}

\author[Y.-F. Wei \& T. Liu]{
Yun-Feng Wei,$^{1,2}$\thanks{E-mail: weiyunfeng@nbu.edu.cn}
and Tong Liu,$^{3}$\thanks{E-mail: tongliu@xmu.edu.cn}
\\
$^{1}$Institute of Fundamental Physics and Quantum Technology, Ningbo University, Ningbo, Zhejiang 315211, China\\
$^{2}$School of Physical Science and Technology, Ningbo University, Ningbo, Zhejiang 315211, China\\
$^{3}$Department of Astronomy, Xiamen University, Xiamen, Fujian 361005, China
}

\date{Accepted XXX. Received YYY; in original form ZZZ}

\pubyear{\the\year{}}

\begin{document}
\label{firstpage}
\pagerange{\pageref{firstpage}--\pageref{lastpage}}
\maketitle

\begin{abstract}

Neutrino-dominated accretion flows (NDAFs) are believed to form during the fallback accretion phase of some core-collapse supernovae (CCSNe). Such systems produce copious neutrino emission, whose cumulative contribution over cosmic history forms the diffuse NDAF neutrino background (DNNB). As neutrinos propagate from the source to Earth, flavour conversion can significantly modify the observed neutrino spectra and consequently the detectability of the DNNB. In this work, based on fallback CCSN simulations, we investigate the effects of progenitor mass, metallicity, and initial explosion energy on neutrino emission from NDAFs. We calculate the heavy-lepton neutrino ($\nu_x$) spectra from NDAFs and incorporate them into DNNB predictions. We find that the unoscillated $\nu_x$ spectra are more than an order of magnitude lower than those of electron antineutrinos $\bar{\nu}_e$. Using the latest neutrino oscillation parameters reported by the Jiangmen Underground Neutrino Observatory (JUNO), we evaluate the impact of flavour conversion on the DNNB and derive the corresponding spectra for both the normal and inverted mass orderings. We further estimate the expected event numbers in JUNO and Hyper-Kamiokande. We find that the predicted DNNB signal is strongly dependent on the neutrino mass ordering. While the DNNB may be detectable in the normal ordering with next-generation neutrino detectors, the signal is significantly suppressed in the inverted ordering, making detection considerably more challenging.

\end{abstract}




\section{Introduction}

Massive stars ($\gtrsim 8 M_{\odot}$) are expected to end their lives as core-collapse supernovae (CCSNe). During the core-collapse process, neutrinos and antineutrinos of all flavours are produced in the core of the newly formed proto-neutron star. These neutrinos carry away most of the gravitational energy released during the collapse \citep[e.g.,][]{Janka2012,Janka2017,Burrows2013}. Eventually, the collapsed core settles into either a neutron star (NS) or a black hole (BH). Over cosmic history, many generations of stars have been born and died. The integrated flux from all past core collapses at cosmological distances constitutes the diffuse supernova neutrino background (DSNB). This background is expected to be detected in the near future and has been extensively studied \citep[see e.g.,][]{Ando2004,Beacom2010,Lunardini2016,Vitagliano2020,Suliga2022}. Meanwhile, fallback accretion is widely believed to occur in CCSNe, especially for more massive progenitors. In the collapsar scenario \citep[e.g.,][]{Woosley1993,MacFadyen1999}, fallback accretion leads to the formation of a BH hyperaccretion disc, which can launch relativistic jets via neutrino annihilation or the Blandford-Znajek (BZ) mechanism \citep{Blandford1977}. If these jets successfully break out of the progenitor envelope, a gamma-ray burst can be triggered. When the mass accretion rate is sufficiently high ($\gtrsim 10^{-3} \,M_{\odot}\,\mathrm{s}^{-1}$), the inner region of the disc becomes extremely dense and hot, and the cooling is dominated by neutrino emission. This disc is known as a neutrino-dominated accretion flow (NDAF), which has been extensively studied in recent decades \citep[see e.g.,][]{Popham1999,Narayan2001,Kohri2002,Lee2005,Gu2006,Chen2007,Janiuk2007,Kawanaka2007,Liu2007,Liu2014,Liu2017,Lei2009,Xue2013,Song2016}.

The cumulative neutrino emission from all NDAFs throughout cosmic history gives rise to the diffuse NDAF neutrino background (DNNB), whose flux may be comparable to that of the DSNB \citep{Nagataki2003,Schilbach2019,Wei2024}. The DNNB also encodes rich information about supernova physics and can provide complementary constraints on the DSNB. The detection of the DNNB would significantly advance our understanding of CCSNe and NDAFs. Moreover, this background can provide valuable insights into neutrino properties. Similar to the DSNB, the DNNB is sensitive to neutrino oscillation physics.

Neutrinos emitted from accretion discs undergo flavour transformation during their propagation to Earth. To calculate the DNNB electron antineutrino ($\bar{\nu}_e$) spectrum after flavour conversion, it is necessary to evaluate both the electron antineutrino and heavy-lepton neutrino components from NDAFs. In \citet{Wei2024}, we studied the effects of the progenitor properties and initial explosion energies on the $\bar{\nu}_e$ spectrum of the DNNB based on CCSN simulations. In this work, to evaluate the impact of neutrino oscillations, we further compute the unoscillated heavy-lepton neutrino spectra from NDAFs and incorporate them into the DNNB prediction, as well as predicting the expected event numbers at Jiangmen Underground Neutrino Observatory (JUNO) and Hyper-Kamiokande (Hyper-K).

The paper is organized as follows. In Section 2, we introduce the NDAF model, the calculation of heavy-lepton neutrino spectra, and the treatment of neutrino flavour conversion in the DNNB, from which the DNNB spectra for both the normal ordering (NO) and inverted ordering (IO) cases are obtained. In Section 3, we evaluate the detectability of the DNNB in JUNO and Hyper-Kamiokande for different mass orderings. Finally, Section 4 presents the discussion and conclusions.

\section{Formulation and models}

\subsection{Formulation of the DNNB}

Accurate predictions of the DNNB require reliable estimates of both the average neutrino emission spectrum and the occurrence rate of NDAFs. Previous studies have shown that the neutrino emission spectra of NDAFs depend sensitively on the properties of their progenitor stars, such as their masses and metallicities \citep{Wei2019,Wei2021}. To account for the diversity of NDAF sources, the initial mass function (IMF)-weighted neutrino spectrum is used in DNNB calculations. The resulting DNNB flux at Earth is given by
\begin{equation}
  \frac{d\Phi }{dE_{\nu}} =c\int_{0}^{\infty } (1+z)\frac{dN}{dE_{\nu} ^{'}}R_{\rm{NDAF}}(z)\left |\frac{dt}{dz} \right |dz,
  \label{eq1}
\end{equation} 
where $E_{\nu}^{'} =E_{\nu}(1+z)$, and $\left |dt/dz \right |=1/H_{0}(1+z)\sqrt{\Omega_{\Lambda}+\Omega _{m}(1+z)^{3}}$. Here, $H_{0}= 70\;\rm{km} \;\rm{s}^{-1}\;\rm{Mpc}^{-1}$, $\Omega _{m}=0.3$, and $\Omega_{\Lambda}=0.7$ are adopted. We set $z_{\rm{max}}=5$, which is large enough to incorporate the majority of the DNNB flux.

Similar to the CCSN rate, the NDAF event rate depends on both the cosmic star formation rate and the IMF, and can be expressed as
\begin{equation}
R_{\rm{NDAF}} (z)=R_{\rm{SFR}}(z)\frac{\int_{M_{\rm{min}}}^{M_{\rm{max}}} \Psi (M)dM}{\int_{0.1}^{125} M\Psi(M)dM},
\label{eq2}
\end{equation} 
where $R_{\rm{SFR}}(z)$ is the cosmic star formation rate in units of $\rm{Mpc}^{-3}\; yr^{-1}$, which can be inferred from observations \citep[e.g.,][]{Hopkins2006,Reddy2008,Rujopakarn2010}. In this work, we adopt the continuous broken power law form proposed by \citet{Yuksel2008},
\begin{equation}
R_{\rm{SFR}} (z)=\dot{\rho} _{0}\left [(1+z)^{\alpha\eta} +(\frac{1+z}{C} )^{\beta \eta} + (\frac{1+z}{D} )^{\gamma \eta}\right]^{1/\eta},
\label{eq3}
\end{equation} 
where $\alpha =3.4$, $\beta =-0.3$, $\gamma =-2$, $\eta =-10$, $C\simeq 5100$, $D\simeq 14$, and $\dot{\rho} _{0}=0.014 ~ \rm{Mpc}^{-3}~ yr^{-1}$. We adopt the Salpeter initial mass function \citep{Salpeter1955}, $\Psi (M)\propto M^{-2.35}$, with a mass range of $0.1-125~M_{\odot}$. 
Here, $M_{\rm{max}}$ and $M_{\rm{min}}$ represent the upper and lower mass limits of progenitors capable of producing NDAFs, respectively. The predicted DNNB is only weakly dependent on $M_{\rm{max}}$ and we fix $M_{\rm{max}}=50\;M_{\odot}$, corresponding to the upper mass limit of progenitors capable of producing NDAFs \citep{Liu2021}. For $M_{\rm{min}}$, we adopt the CCSN simulation results in \citet{Wei2024}, which are summarized in Table \ref{tab1}. We adopt the piston approach to carry out spherically symmetric explosion simulations \citep{Woosley1995,Woosley2002}, in which the initial explosion conditions at the inner boundary are determined by the motion of the piston. A unidirectional outflowing inner boundary condition is set at $r=10^9$ cm. To mimic the outward blast passing through the inner boundary, additional energy is injected into the innermost cell adjacent to the inner boundary over a short timescale. The injected energy defines the initial explosion energy, for which we adopt three representative values of 2, 4, and $8\,B$ ($1\,B=10^{51}~\rm erg$).  For more details on the simulation, the reader is referred to \citet{Liu2021a} and \citet{Wei2021}.

We record the evolution of the fallback mass supply rate at the inner boundary. Ignoring the disc outflows, we roughly consider the mass supply rate as the mass accretion rate of the disc. The value of $M_{\rm{min}}$ is determined by whether the initial mass accretion rate reaches the NDAF ignition threshold ($\gtrsim 10^{-3} \,M_{\odot}\,\mathrm{s}^{-1}$). This initial accretion rate depends jointly on the presupernova density structure and the initial explosion energy \citep{Wei2021}. For a given explosion energy, a more compact presupernova structure generally leads to stronger fallback accretion. In our simulations, $M_{\rm{min}}$ is obtained by searching the minimum progenitor mass that produces a BH and satisfies the NDAF criterion. Because the presupernova structure is affected by the metallicity, the resulting value of $M_{\min}$ is metallicity dependent. At a fixed metallicity, a lower explosion energy ejects less stellar material and enhances fallback accretion, allowing lower-mass progenitors to satisfy the NDAF formation criterion and thereby reducing $M_{\min}$. Thus, $M_{\min}$ is jointly determined by the progenitor metallicity and explosion energy. Furthermore, assuming that the minimum and maximum initial masses of core-collapse progenitors are 8 $M_\odot$ and 125 $M_\odot$, respectively, we estimate the fraction of core-collapse events that produce NDAFs ($f_{\rm NDAF}$). The corresponding results are presented in the last column of Table \ref{tab1}. Depending on metallicity and explosion energy, the NDAF progenitor fraction ranges from several percent to about twenty percent of all core-collapse progenitors.

In principle, the metallicities of massive-star progenitors follow a distribution that evolves with redshift. A complete calculation would therefore require the stellar IMF to be convolved with a redshift-dependent metallicity distribution. Because such a treatment introduces substantial additional uncertainties, we do not explicitly account for cosmic metallicity evolution in this work. Instead, we consider three representative metallicities separately, assuming that all progenitors have the same metallicity in each calculation. Similar to our treatment of the explosion energy, this approach is intended to isolate the effect of metallicity on the NDAF event rate and the resulting DNNB. Consequently, the results for the three metallicities represent separate illustrative scenarios rather than a metallicity-averaged cosmic prediction. For each metallicity and explosion energy, only progenitors within the NDAF-forming mass range determined from our simulations are included; progenitors that leave NS remnants or otherwise fail to satisfy the NDAF formation criterion are excluded.

\begin{table} \centering \caption{The minimum progenitor mass, $M_{\rm min}$, and the fraction of core-collapse events producing NDAFs, $f_{\rm NDAF}$, for different metallicities and initial explosion energies.} \label{tab1} \begin{tabular}{cccc} 
\hline 
Metallicity & Initial explosion energy & $M_{\rm min}$ & $f_{\rm NDAF}$ \\ ($Z/Z_\odot$) & (B) & ($M_\odot$) & (\%) \\ 
\hline 
0 & 2 & 30 & 8.6 \\ 
0.01 & 2 & 20 & 21.1 \\ 
1 & 2 & 30 & 8.6 \\ 
0.01 & 4 & 20 & 21.1 \\ 
0.01 & 8 & 40 & 3.0 \\ \hline \end{tabular} \end{table}

\subsection{Neutrinos from NDAFs}

The inner regions of NDAFs are extremely dense and hot, with electrons in a highly degenerate state. Neutrino cooling is efficient in the disc, primarily dominated by Urca processes, with additional contributions from electron-positron pair annihilation, plasma decay, and nucleon-nucleon bremsstrahlung \citep{Liu2016}. The Urca processes predominantly generate electron neutrinos and antineutrinos, while heavy-lepton neutrinos are mainly produced via pair annihilation and nucleon-nucleon bremsstrahlung. As a result, the initial neutrino spectra are flavour dependent. Here, we collectively denote $\nu_\mu$, $\bar{\nu}_\mu$, $\nu_\tau$, and $\bar{\nu}_\tau$ as $\nu_x$, since the differences in their emission processes are subtle.

The generic models of the neutrino spectrum for NDAFs are adopted from the results of CCSN numerical simulations. In \cite{Wei2024}, we carried out a series of core-collapse supernova simulations using the Athena++ code. These simulations provide the time-dependent properties of the central hyperaccreting BH systems, including the BH mass, spin, and accretion rate, which are subsequently used to calculate electron antineutrino spectra of NDAFs. Here, we further calculate the unoscillated heavy-lepton neutrino spectra from NDAFs. Based on the global NDAF solutions of \citet{Xue2013}, we obtain the fitting formulae for the cooling rate due to heavy-lepton neutrino emission and temperature of the disc as a function of the BH mass and spin, the mass accretion rate and radius, i.e.,
\begin{align}
\log Q_{{\nu}_{x}}\;(\rm{erg}\;\rm{cm}^{-2}\;s^{-1}) =\ &39.52-0.25m_{\rm{BH}}+0.88a_{*}
\notag
\\&+1.84\log \dot{m}-3.78\log r,
\label{eq4}
\end{align}
and
\begin{align}
\log T \;(\rm{K})=\ &11.23-0.04m_{\rm{BH}}+0.10a_{*}+0.23\log \dot{m}
\notag
\\&-0.86\log r,
\label{eq5}
\end{align}
where $m_{\rm{BH}}=M_{\rm{BH}}/M_{\odot }$, $\dot{m}=\dot{M}/M_\odot~\rm s^{-1}$, and $r=R/R_g$ are the dimensionless BH mass, accretion rate, and radius, respectively. Here, $R_g=2GM_{\rm{BH}}/c^2$ is the Schwarzschild radius. The local heavy-lepton neutrino spectra are assumed to follow a Fermi--Dirac distribution and are normalised by the cooling rate $Q_{{\nu}_{x}}$.

\begin{figure}
\centering
\includegraphics[angle=0,scale=0.32]{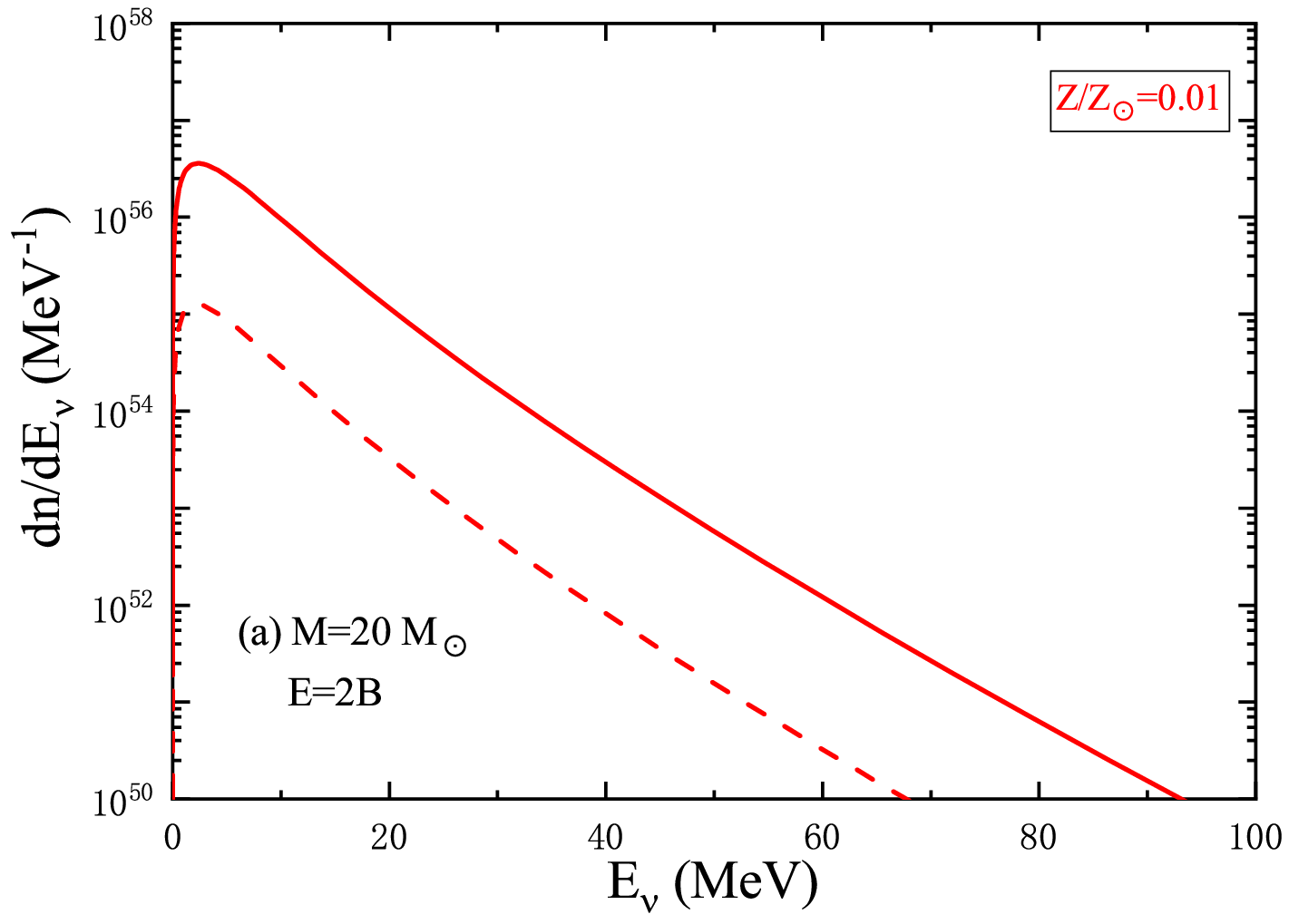}
\includegraphics[angle=0,scale=0.32]{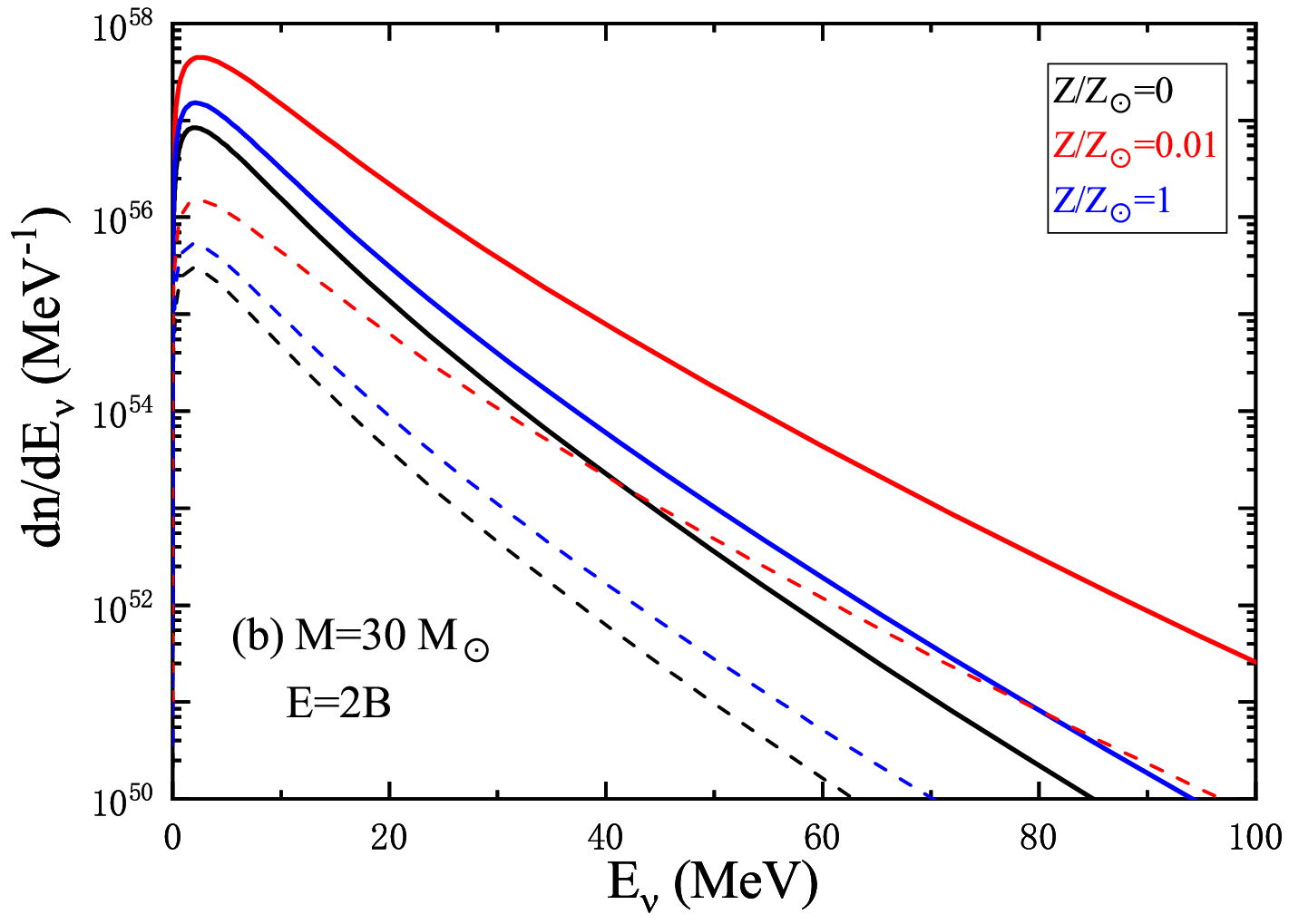}
\includegraphics[angle=0,scale=0.32]{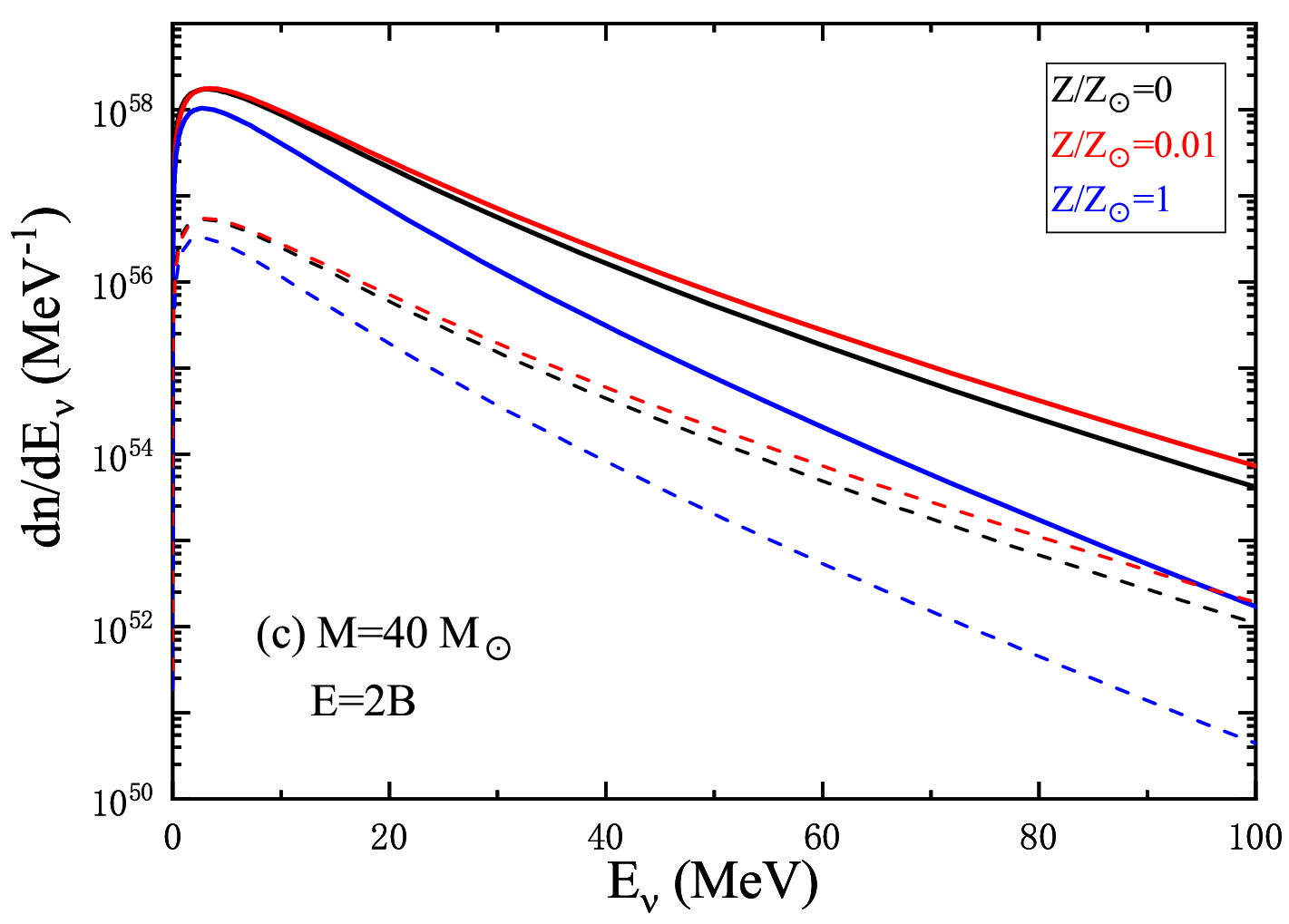}
\caption{Time-integrated unoscillated antineutrino spectra of NDAFs with different masses and metallicities of progenitors. The black, red, and blue curves correspond to progenitor star metallicities of $Z/Z_{\odot}=0$, 0.01, and 1, respectively. The solid lines correspond to $\bar{\nu}_{\rm{e}}$, while the dashed lines represent $\nu_x$. The initial explosion energy is $2\,B$.}
\label{fig1}
\end{figure}

Using the NDAF evolutionary data derived by \citet{Wei2024}, we calculate the time-integrated heavy-lepton neutrino spectra for NDAFs formed from progenitors with different masses and metallicities. The results are shown in Fig.~\ref{fig1}. Further details of the neutrino spectral calculations can be found in \citet{Wei2024}. Panels (a), (b), and (c) correspond to progenitors with masses of 20, 30, and $40\,M_{\odot}$, respectively. The initial explosion energy is set to $2\,B$. The solid lines represent the $\bar{\nu}_e$ spectra adopted from \citet{Wei2024}, while the dashed lines denote the $\nu_x$ spectra calculated in this work. The dependence of the $\nu_x$ spectra on progenitor properties is similar to that of $\bar{\nu}_e$. It can be seen that the $\nu_x$ spectrum is more than an order of magnitude lower than that of $\bar{\nu}_e$. This reflects the dominance of Urca processes in NDAFs, which preferentially produce electron neutrinos and antineutrinos. In Fig.~\ref{fig1}(a), we present the neutrino spectra of NDAFs only for progenitors with metallicity $Z/Z_{\odot} = 0.01$, since in our simulations progenitors with metallicities of $Z/Z_{\odot} = 0$ and 1 form NSs rather than BHs. In general, the antineutrino spectra for solar-metallicity progenitors are significantly lower than those for low-metallicity cases. This can be attributed to the lower core compactness at solar metallicity, which results in a weaker initial fallback accretion.

\begin{figure}
\centering
\includegraphics[angle=0,scale=0.32]{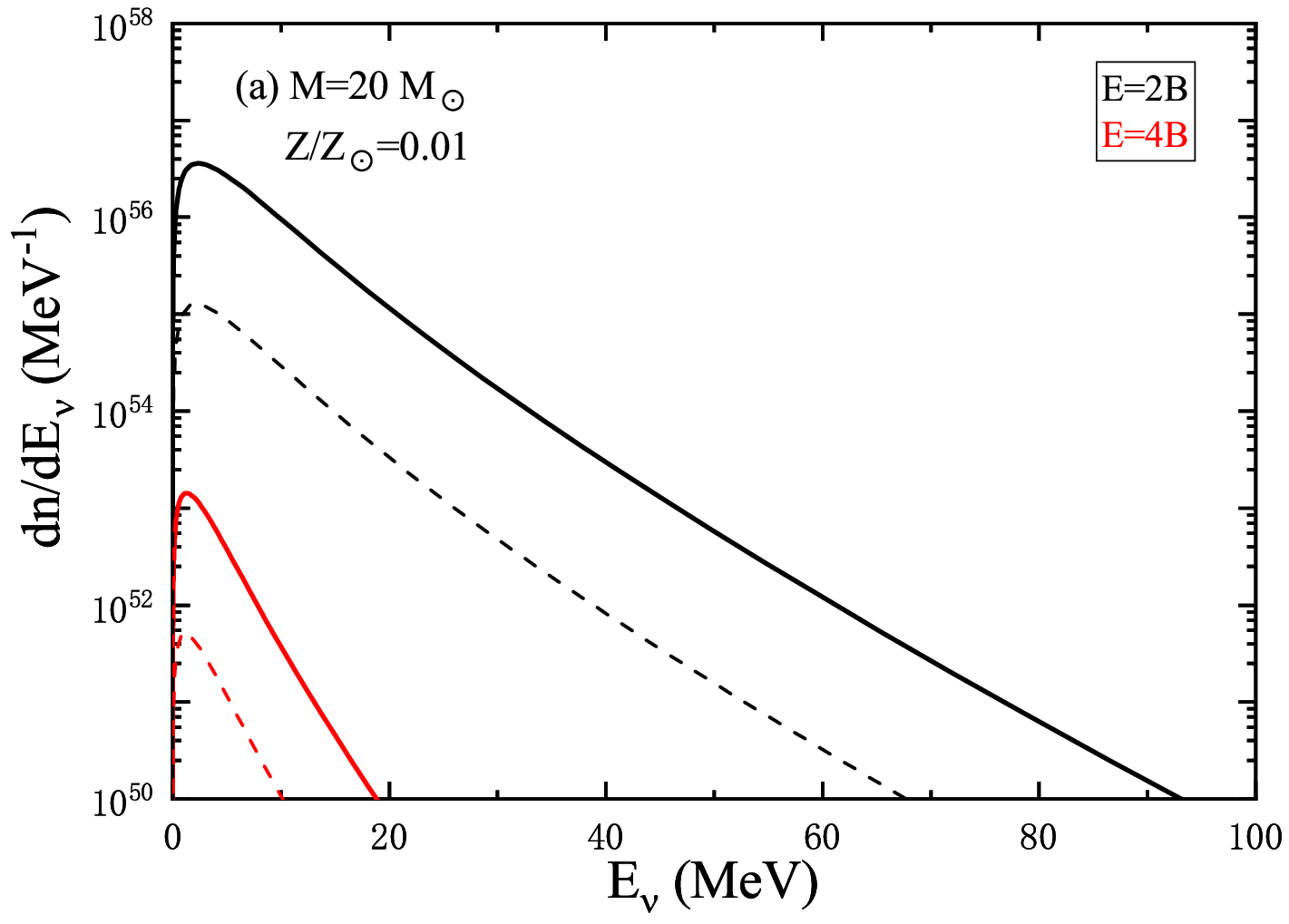}
\includegraphics[angle=0,scale=0.32]{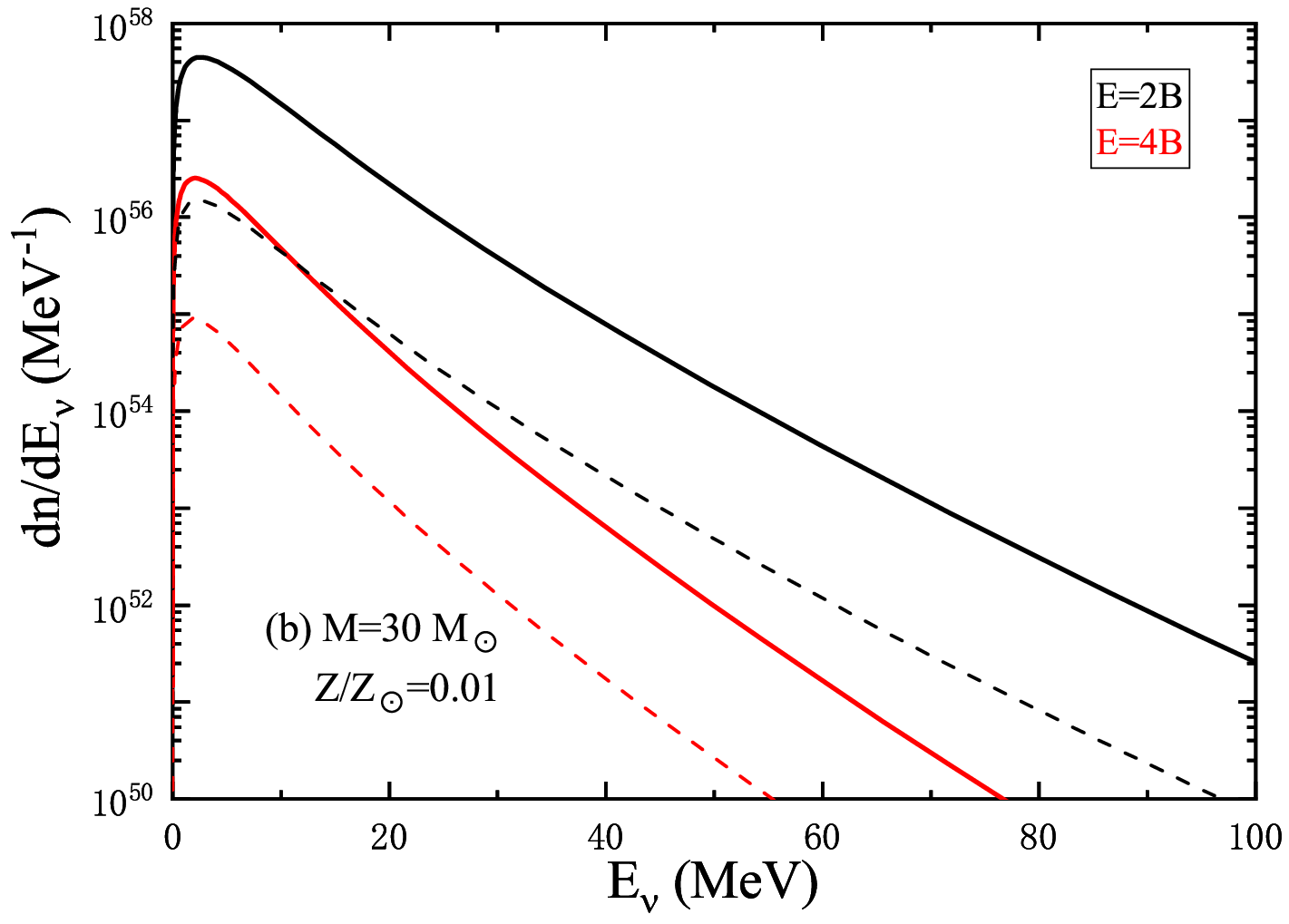}
\includegraphics[angle=0,scale=0.32]{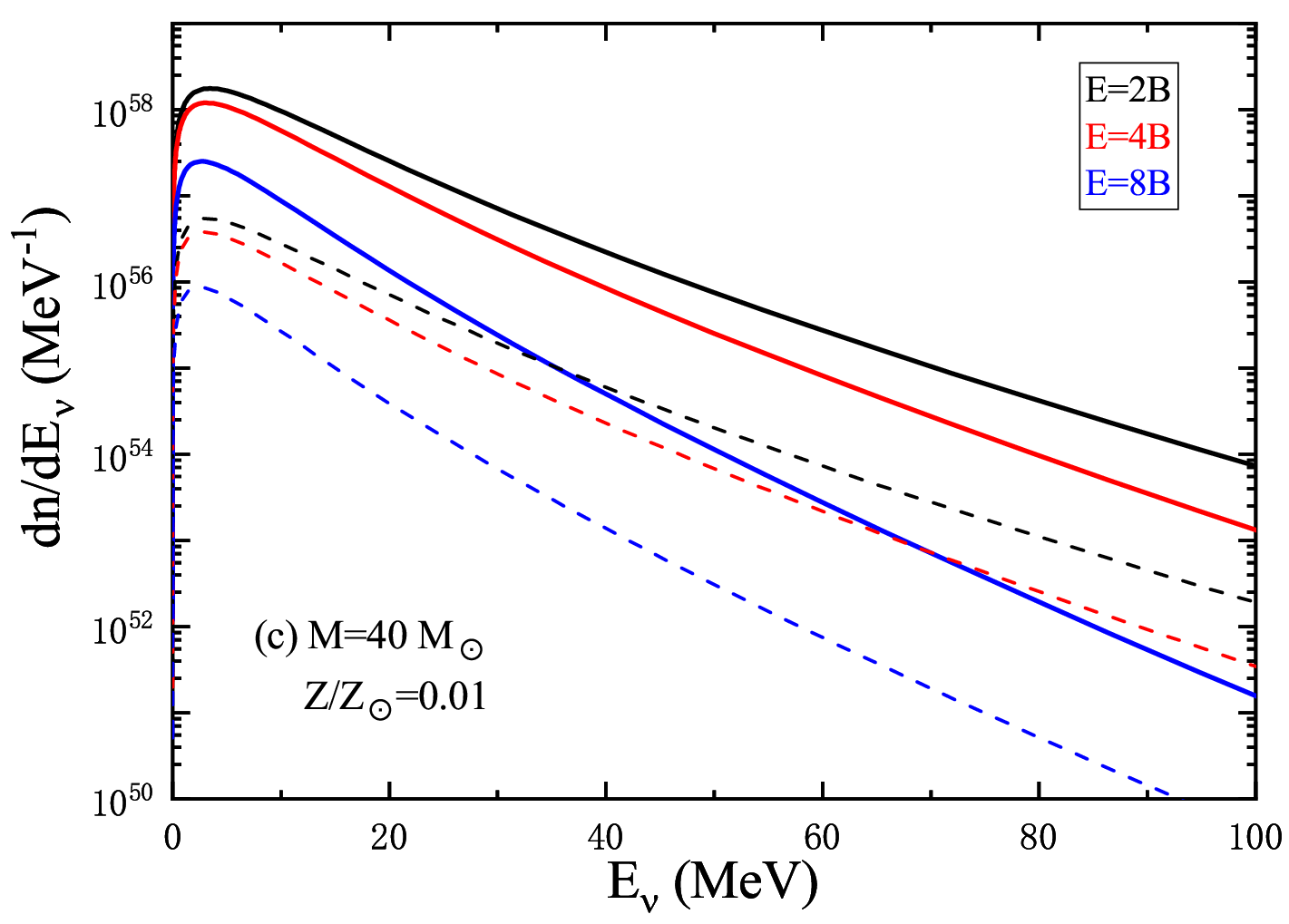}
\caption{Time-integrated unoscillated antineutrino spectra of NDAFs for different initial explosion energies. The black, red, and blue curves correspond to initial explosion energies of 2, 4, and $8\,B$, respectively. The solid lines correspond to $\bar{\nu}_{\rm{e}}$, while the dashed lines represent $\nu_x$. The metallicity is set as $Z/Z_{\odot}=0.01$.}
\label{fig2}
\end{figure}

In Fig.~\ref{fig2}, we present the effects of the initial explosion energy on the unoscillated antineutrino spectra of NDAFs, with the metallicity fixed at $Z/Z_{\odot} = 0.01$. The black, red, and blue curves correspond to initial explosion energies of 2, 4, and $8\,B$, respectively. As the explosion energy decreases, the spectral amplitudes increase, reflecting the stronger fallback accretion associated with weaker explosions. In Fig.~\ref{fig2} (a) and Fig.~\ref{fig2} (b), the $8\,B$ case is not shown because these progenitors form NSs.

Then the IMF-weighted neutrino spectrum for a population of progenitors can be computed by weighting each progenitor according to the initial mass function, i.e.,
\begin{equation}
\frac{dN}{dE_{\nu}} =\sum_{i}\frac{\int_{\bigtriangleup M _{i}} \Psi(M)dM}{\int_{M_{\rm{min}}}^{M_{\rm{max}}} \Psi (M)dM} F_{i} (E_{\nu }),
\label{eq6}
\end{equation}
where $\bigtriangleup M _{i}$ is the mass range of mass bin $i$, and $F_{i} (E_{\nu })$ is the neutrino spectrum of an NDAF corresponding to a progenitor of mass $M_{i}$.

\subsection{Neutrino oscillations}

\begin{figure}
\centering
\includegraphics[angle=0,scale=0.32]{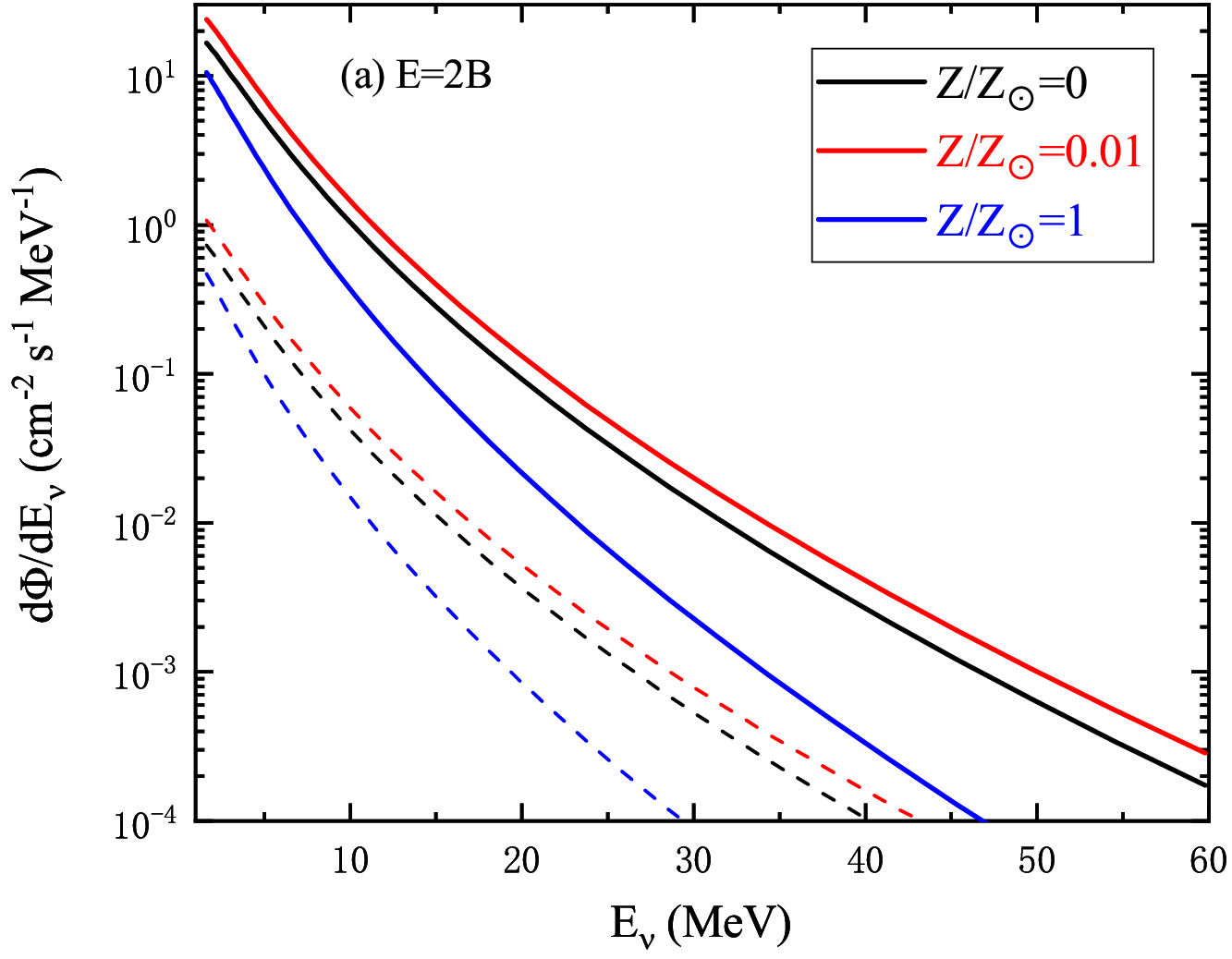}
\includegraphics[angle=0,scale=0.32]{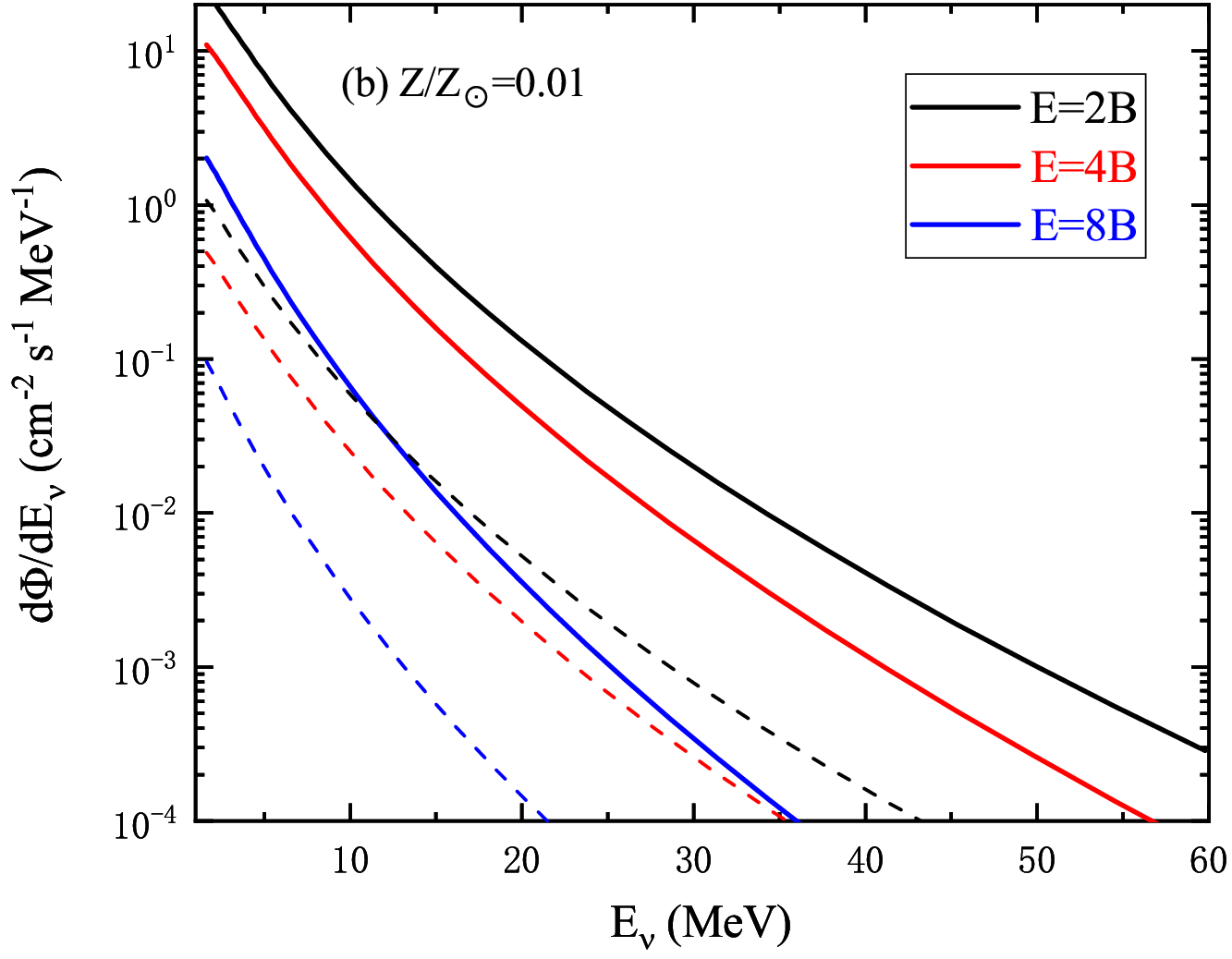}
\caption{The $\bar{\nu}_e$ spectra of the DNNB as a function of neutrino energy. Solid (dashed) lines correspond to the normal ordering (NO) and inverted ordering (IO). Panels (a) and (b) show the effects of progenitor metallicity and initial explosion energy on the DNNB, respectively.}
\label{fig3}
\end{figure}

Since neutrino flavour eigenstates are not identical to mass eigenstates, neutrinos undergo flavour mixing during propagation. The flavour eigenstates $\nu_e$, $\nu_\mu$, $\nu_\tau$ are related to the mass eigenstates by the unitary mixing matrix as
\begin{equation}
    \nu_\alpha = \sum_i U_{\alpha i} \, \nu_i,
    \label{eq7}
\end{equation}
and the mixing matrix U can be expressed as
\begin{equation}
\begin{aligned}
U& = \\
& \begin{pmatrix}
c_{12} c_{13} & s_{12} c_{13} & s_{13} e^{-i\delta} \\
- s_{12} c_{23} - c_{12} s_{23} s_{13} e^{i\delta}
& c_{12} c_{23} - s_{12} s_{23} s_{13} e^{i\delta}
& s_{23} c_{13} \\
s_{12} s_{23} - c_{12} c_{23} s_{13} e^{i\delta}
& - c_{12} s_{23} - s_{12} c_{23} s_{13} e^{i\delta}
& c_{23} c_{13}
\end{pmatrix},
\end{aligned}
\label{eq8}
\end{equation}
where $c_{ij} \equiv \cos\theta_{ij}$ and $s_{ij} \equiv \sin\theta_{ij}$, and $\delta$ is the CP-violating phase.

Neutrinos emitted from NDAFs undergo oscillations during their propagation to Earth. When passing through a collapsing star, neutrinos (and antineutrinos) experience both collective and matter-induced (MSW) flavour conversions \citep{Wolfenstein1978,Mikheev1986,Duan2006,Duan2010,Mirizzi2016}. In this work, we neglect the effects of collective neutrino oscillations. When produced in the dense NDAF environment, flavour states are approximately locked to matter eigenstates owing to the large matter potentials. Here, we consider two different mass orderings: normal mass ordering (NO) ($m_1 < m_2 < m_3$) and inverted mass ordering (IO) ($m_3 < m_1 < m_2$) \citep{Esteban2020}. Then, for the NO, $\bar{\nu}_e$ exit as $\bar{\nu}_1$, while $\bar{\nu}_x$ exit as $\bar{\nu}_2$ and $\bar{\nu}_3$. For the IO,  $\bar{\nu}_e$ predominantly emerge as $\bar{\nu}_3$, while $\bar{\nu}_x$ emerge as $\bar{\nu}_1$ and $\bar{\nu}_2$  \citep{Dighe2000}. Assuming the adiabatic flavour conversion, the DNNB flux spectrum of $\bar{\nu}_e$ after including the effect of flavour conversions can be expressed as
\begin{equation}
\frac{d\Phi_{\bar{\nu}_e}}{dE_{\nu}}
=
\bar{p}\frac{d\Phi^{0}_{\bar{\nu}_e}}{dE_{\nu}}
+
(1-\bar{p})\frac{d\Phi^{0}_{\nu_x}}{dE_{\nu}},
\label{eq9}
\end{equation}
where $d\Phi^{0}_{\bar{\nu}_e}/dE_{\nu}$ and $d\Phi^{0}_{\nu_x}/dE_{\nu}$ are the unoscillated spectra for $\bar{\nu}_e$ and $\nu_x$. Possible non-adiabatic effects induced by shock wave propagation \citep{Galais2010} or sharp density discontinuities are beyond the scope of this work. $\bar{p}$ is the survival probability of $\bar{\nu}_e$. Under the adiabatic MSW approximation, $\bar{p}\simeq \left|U_{e1}\right|^2$ for the NO and $\bar{p}\simeq \left|U_{e3}\right|^2$ for the IO \citep{Dighe2000}. We adopt the latest neutrino oscillation parameters measured by JUNO, which is a 20 kton liquid-scintillator detector located 52.5 km from multiple reactor cores, designed to precisely measure neutrino oscillation parameters and determine the neutrino mass ordering through reactor antineutrino oscillations \citep{Juno2022}. In this work, we use $\sin^2\theta_{12} = 0.3092$ \citep{Abusleme2025}. The $\theta_{13}$ is small and we adopt $\sin^2\theta_{13} = 0.0223$ \citep{Workman2022}.

\begin{figure}
\centering
\includegraphics[angle=0,scale=0.35]{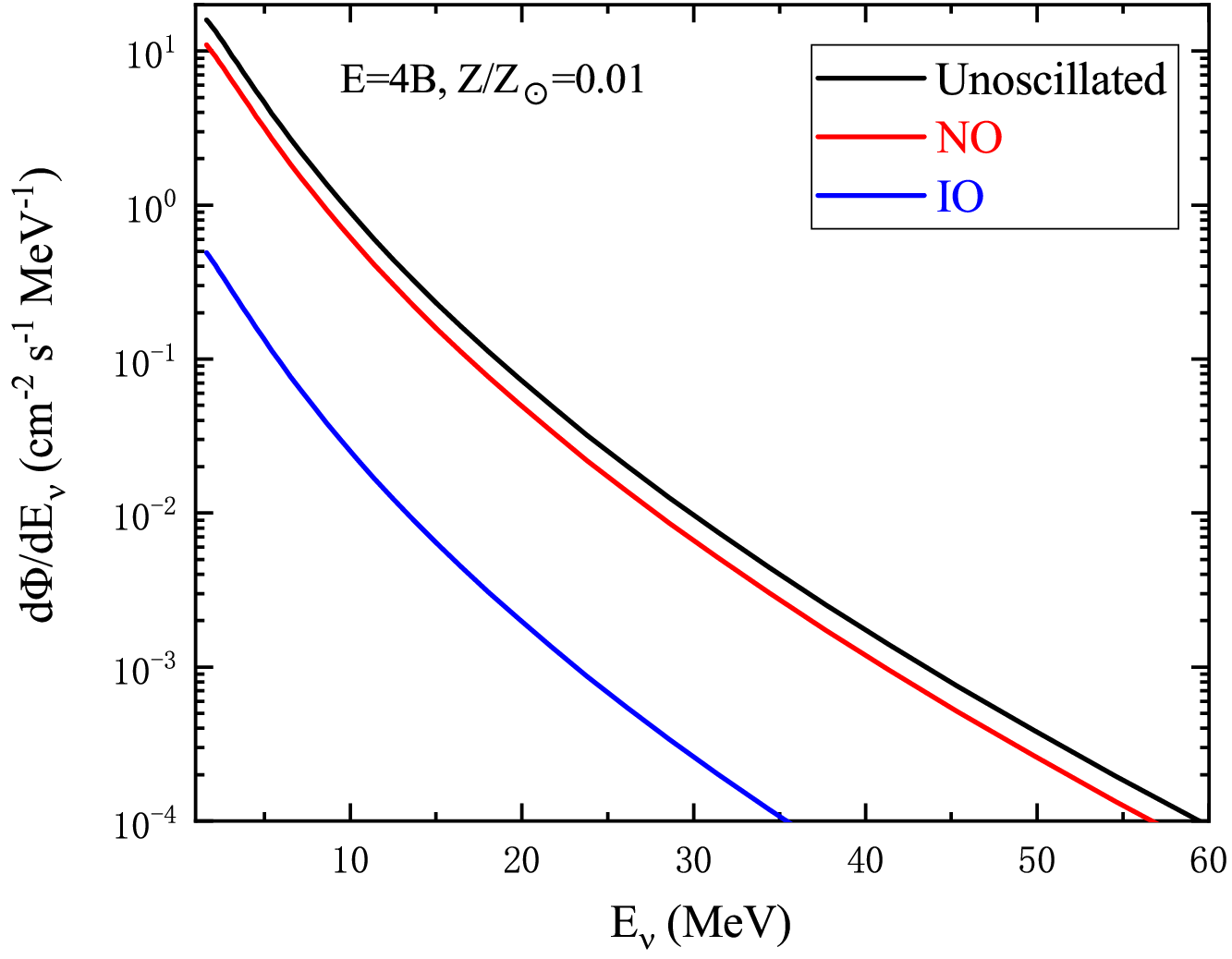}
\caption{The $\bar{\nu}_e$ spectra of the DNNB for the unoscillated, NO, and IO cases with $E=4B$ and $Z/Z_{\odot}=0.01$. The black, red, and blue lines denote the unoscillated, NO, and IO spectra, respectively.}
\label{fig4}
\end{figure}

\begin{figure*}
\centering
\includegraphics[angle=0,scale=0.32]{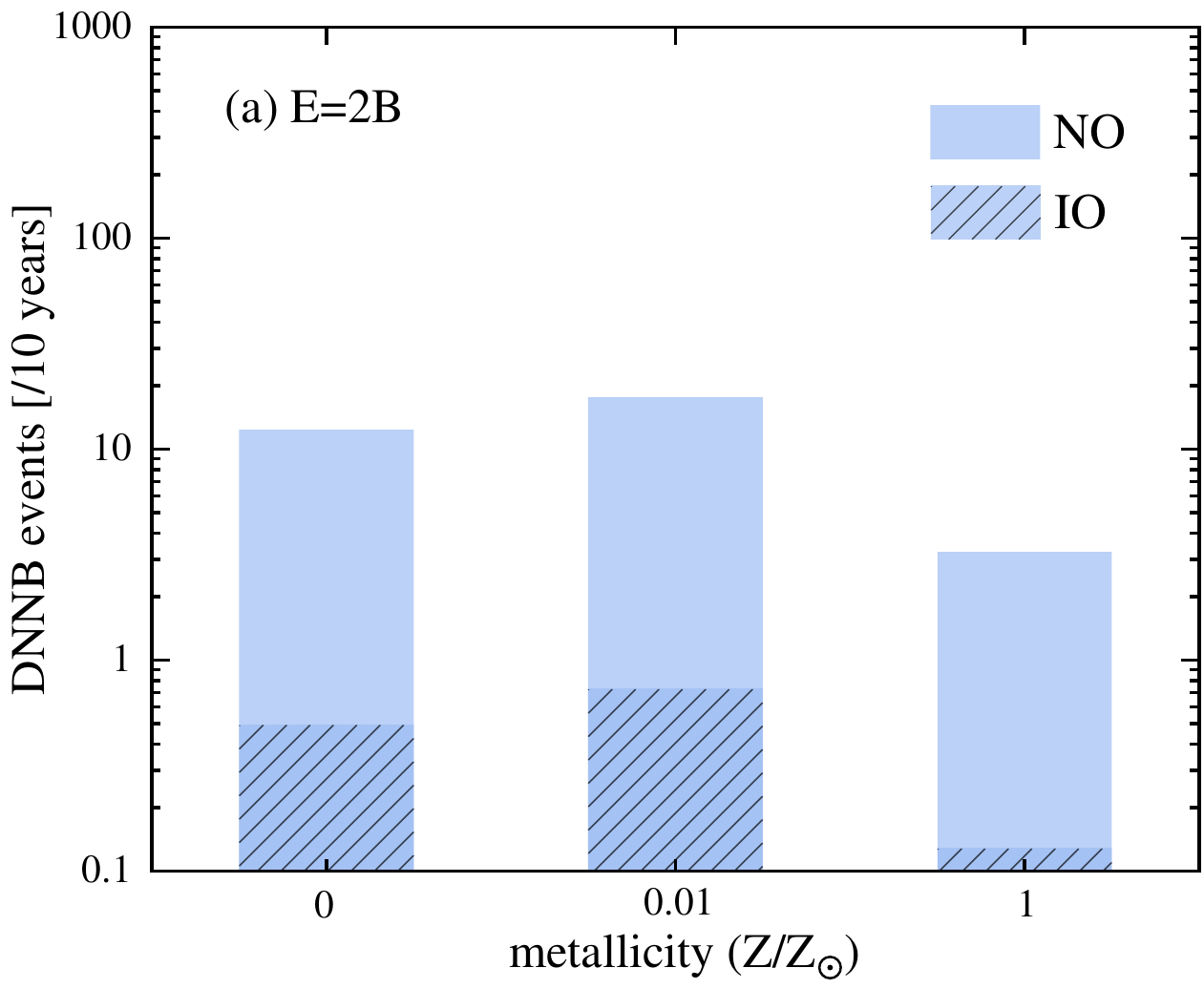}
\includegraphics[angle=0,scale=0.32]{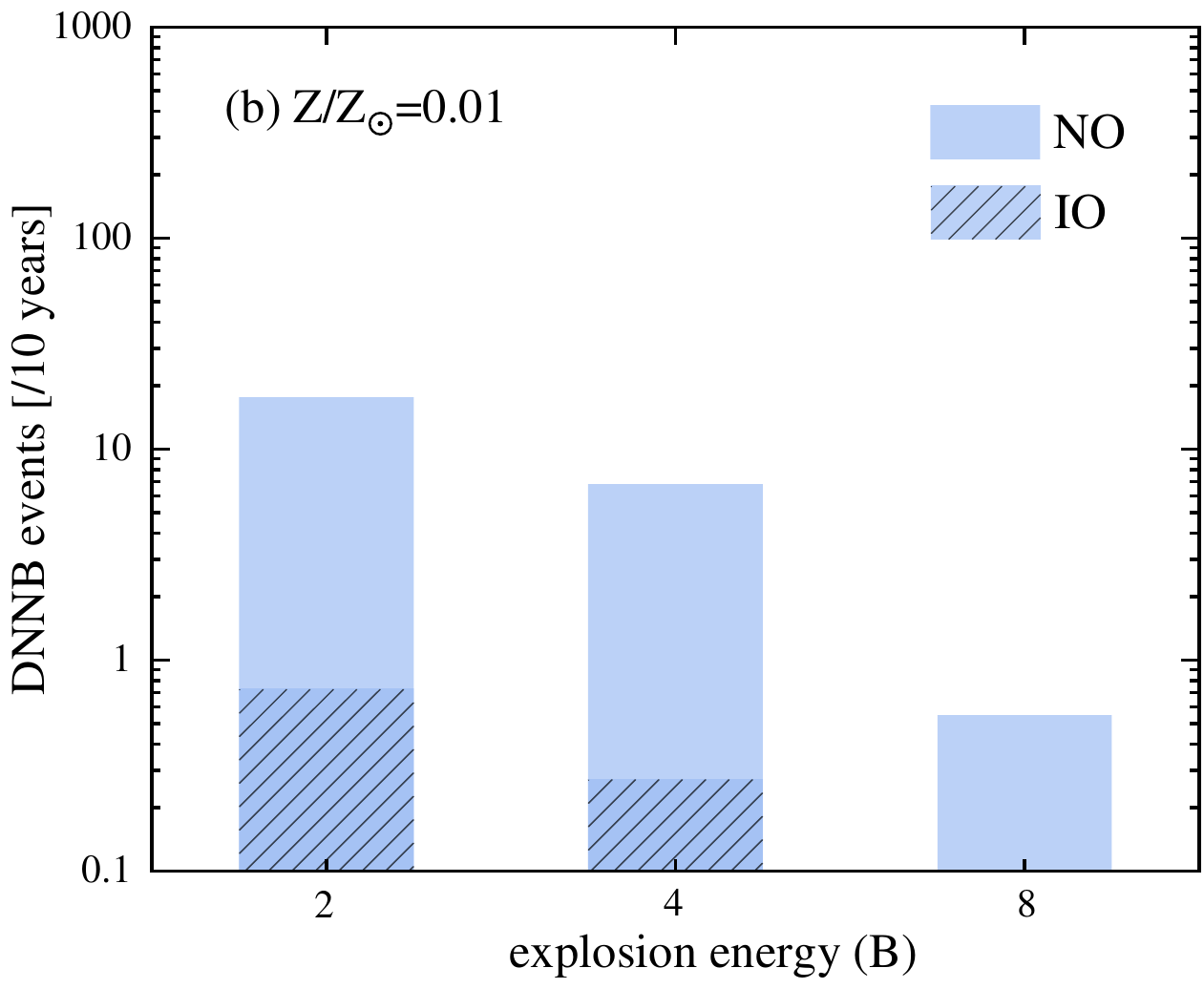}
\caption{Predicted DNNB event numbers in JUNO over a 10-year observation period. The left and right panels show the effects of progenitor metallicity and initial explosion energy, respectively. Solid bars denote the NO, whereas hatched bars indicate the IO.}
\label{fig5}
\end{figure*}

\begin{figure*}
\centering
\includegraphics[angle=0,scale=0.32]{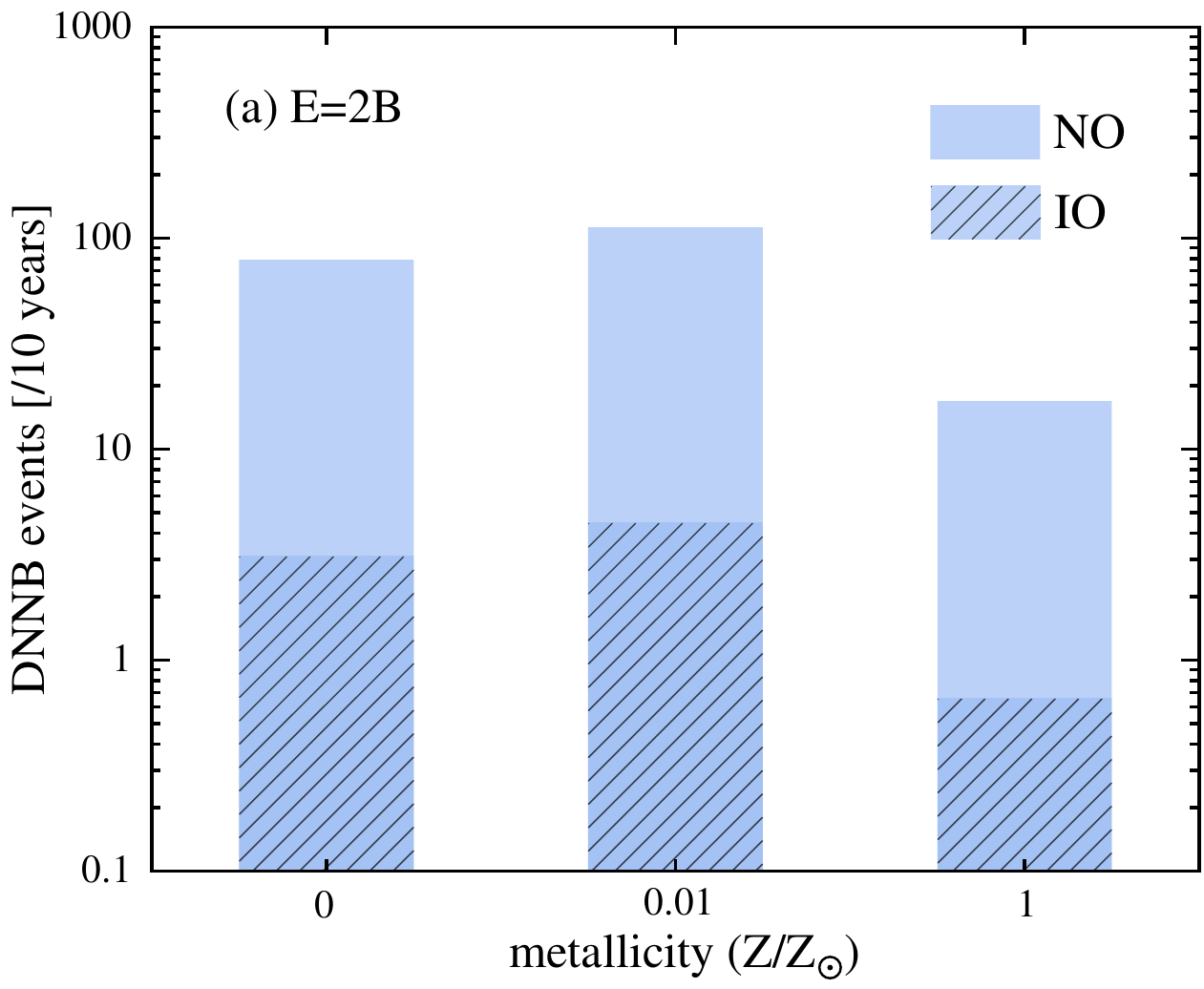}
\includegraphics[angle=0,scale=0.32]{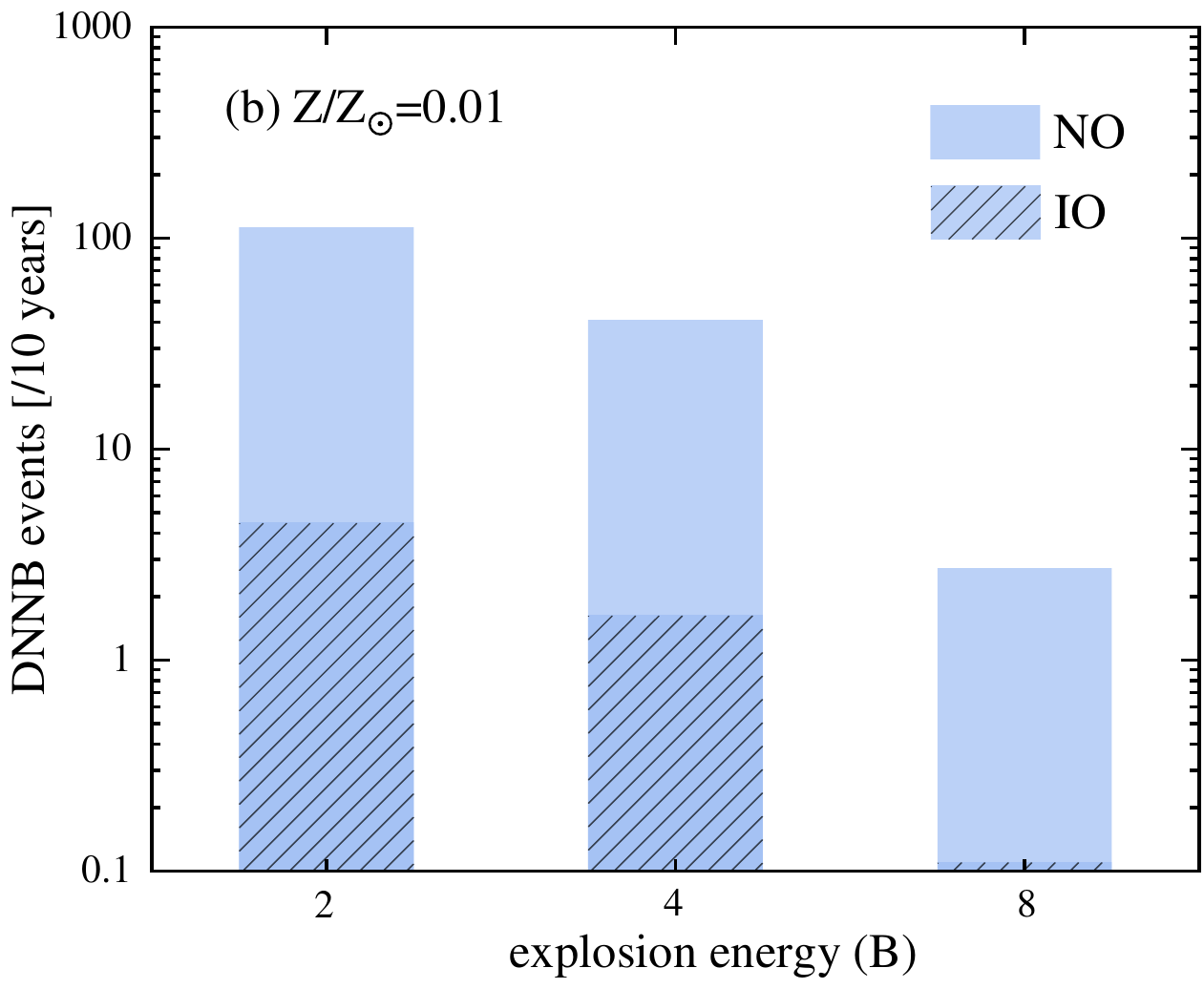}
\caption{Same as Fig.~\ref{fig5}, but for Hyper-K.}
\label{fig6}
\end{figure*}

Fig.~\ref{fig3} illustrates the effects of the metallicity and the initial explosion energy on the DNNB flux spectrum. In Fig.~\ref{fig3} (a), the initial explosion energy of all progenitors is set as  $2\,B$. The black, red, and blue curves correspond to $Z/Z_{\odot}=0$, $0.01$, and $1$, respectively. In Fig.~\ref{fig3} (b), we assume that all progenitors have the same metallicity, ($Z/Z_{\odot}=0.01$). The black, red, and blue curves correspond to the initial explosion energy of 2, 4, and $8\,B$, respectively. A weaker explosion energy leads to stronger fallback accretion, which in turn enhances the neutrino emission and the event rate of NDAFs. In both panels, the solid lines represent the NO, while the dashed lines represent the IO. To illustrate the impact of neutrino flavour conversion, we show the $\bar{\nu}_e$ spectra of the DNNB for the unoscillated, NO, and IO cases in Fig.~\ref{fig4}. The black, red, and blue curves correspond to the unoscillated, NO, and IO cases, respectively. Neutrino flavour conversion significantly modifies the flux due to the large difference between the primary  $\bar{\nu}_e$ and $\nu_x$ spectra, with the IO case showing a stronger suppression.

\section{Detection rates}

The detection rate of the DNNB neutrinos is calculated via the inverse beta decay (IBD) reaction, $\bar{\nu}_e + p \rightarrow n + e^+$. The event rate can be written as
\begin{equation}
\frac{dN_{\rm e^+} }{dE_{\rm e^+} }(E_{\rm e^+}) =N_{\rm{t}} \sigma (E_{\bar{\nu}_{e}}) \frac{d\Phi_{\bar{\nu}_e}}{dE_{\nu}},
\label{eq10}
\end{equation}
where $\sigma(E_\nu)$ is the IBD cross section \citep{Vogel1999,Strumia2003}. The positron energy is expressed as $E_{\rm e^+}= E_{\bar{\nu}_{e}} -\Delta c^{2}$, where $\Delta$ is the neutron-proton mass difference. Here, $N_{\rm t}$ denotes the number of free protons in the detector.

Then, the total number of detected events is obtained by integrating the event-rate spectrum over the analysis energy window $[E_{\rm min}, E_{\rm max}]$ and multiplying by the observation time $T_{\rm obs}$, i.e.,
\begin{equation}
N_{\rm det}
=
\epsilon_{\rm sig} T_{\rm obs}
\int_{E_{\rm min}}^{E_{\rm max}}
\frac{dN_{\rm event}}{dE}\, dE,
\label{eq11}
\end{equation}
where $\epsilon_{\rm sig}$ is the signal efficiency at the detector. The analysis energy window for DNNB is selected to avoid the dominant low-energy backgrounds, such as reactor and solar neutrinos, as well as the high-energy background from atmospheric neutrinos.

For detection, we consider the JUNO (liquid-scintillator detector) and Hyper-K (water-Cherenkov detector), with 20 and 374 kton inner volumes, respectively. For the optimistic scenarios, corresponding to low initial explosion energies, the predicted DNNB flux can be comparable to that of the DSNB. Super-Kamiokande (Super-K) has searched for the DSNB, but no signal has been detected so far. The current upper limit corresponds to fewer than two events in the energy range of $18-26$ MeV over 1496 days of observation \citep{Horiuchi2009}. Compared with Super-K, JUNO has a comparable number of target protons ($N_{\rm t} = 1.45\times10^{33}$), while offering a lower energy threshold and higher energy resolution. Following \citet{Li2022}, we adopt an overall signal efficiency of $\epsilon_{\rm sig}=0.5$ and an analysis energy window of $12-30$ MeV for JUNO. Hyper-K is an upcoming water Cherenkov detector and the successor to Super-K. It consists of two cylindrical tanks, each with a fiducial mass of 187 kton, corresponding to a total number of target protons of $N_{\rm t} = 2.5 \times10^{34}$ \citep{Abe2011}. We adopt a signal efficiency of $\epsilon_{\rm sig}=0.67$ and an analysis energy window of $18-26$ MeV \citep{Muller2018}.

Fig.~\ref{fig5} and Fig.~\ref{fig6} present the expected DNNB event numbers in JUNO and Hyper-K, respectively, for a 10-year observation period. Solid bars denote the NO, whereas hatched bars indicate the IO. The effects of progenitor metallicity and initial explosion energy on the DNNB event numbers are consistent with those on the DNNB flux. For solar-metallicity progenitors and energetic explosions, the expected DNNB event numbers are substantially reduced, rendering the DNNB nearly undetectable. The neutrino mass ordering has a significant impact on the detectability of the DNNB. In all models, the predicted event numbers in the NO case exceed those in the IO case by more than an order of magnitude. Consequently, the DNNB may be detectable in the most optimistic NO scenarios, particularly for low-metallicity progenitors and weak explosions. In contrast, the predicted event numbers in the IO case are generally too low for a statistically significant detection.

Hyper-K offers the most promising prospects for DNNB detection owing to its enormous fiducial mass. For the most optimistic models, the predicted event numbers can exceed one hundred events over a decade, indicating a realistic possibility of detection. Although JUNO is expected to observe lower statistics than Hyper-K, its wider energy coverage and high energy resolution make it a valuable complementary detector for probing the DNNB spectral shape and enhancing the overall sensitivity in joint analyses. Moreover, JUNO may determine the neutrino mass ordering after several years of operation. Given the strong dependence of the DNNB signal on the mass ordering, such a measurement would substantially reduce the uncertainty in DNNB predictions and improve the reliability of detectability estimates.

\section{Conclusions and Discussion}

We have investigated the impact of neutrino flavour conversion on the DNNB spectra. For the first time, we systematically evaluate the effects of progenitor mass and metallicity on the unoscillated $\nu_x$ spectra from NDAFs. The $\nu_x$ spectra are lower than those of $\bar{\nu}_e$ by more than an order of magnitude. Based on these results, we calculate the DNNB spectra under both normal and inverted mass orderings and predict the corresponding event numbers at JUNO and Hyper-K. Both the progenitor mass and metallicity can affect the neutrino spectra and the event rates of NDAFs. In particular, lower initial explosion energies lead to stronger fallback accretion, thereby enhancing the neutrino emission and increasing the detectability of the DNNB. The strong dependence of the DNNB signal on neutrino flavour conversion further implies that future MeV neutrino observations may serve as a complementary probe of the neutrino mass ordering. For the NO, flavour conversion leads to a larger survival fraction of the original $\bar{\nu}_e$ component, making the DNNB potentially detectable, particularly with Hyper-K and, in the most optimistic cases, with JUNO over decade-long observations. In contrast, for the IO, the $\bar{\nu}_e$ flux is suppressed, leading to a significantly lower expected event rate and making the DNNB difficult to detect.

From an observational perspective, an important issue is how to distinguish the DNNB from the conventional DSNB. Because both predictions still contain large uncertainties, a precise separation is difficult. In our calculations, the DNNB can approach the DSNB level only for the NO, whereas it remains significantly below the DSNB for the IO. One possible signature is the high-energy spectrum: if weak explosions are common, NDAFs may produce harder neutrino spectra, causing the DNNB to decline more slowly than the DSNB at high energies. Additional information may be provided by multi-messenger observations. In \citet{Wei2024a}, we investigated the stochastic gravitational wave background (SGWB) from NDAFs. A future detection or meaningful upper limit on this background could constrain the cosmic NDAF event rate and, when combined with diffuse neutrino observations, the average neutrino emission per NDAF event. Such joint constraints would help assess the NDAF contribution to the observed diffuse neutrino background and thereby improve the separation of the DNNB from the DSNB.

Note that self-induced (collective) neutrino flavour transformations can also lead to flavour exchange between $\bar{\nu}_e$ and $\nu_x$ \citep{Duan2006,Duan2010}. However, the development of collective oscillations in dense accretion disc environments remains highly uncertain due to multi-angle matter suppression and the lack of a self-consistent treatment of neutrino angular distributions. In this work, we therefore neglect the effect of neutrino self-induced oscillations on the DNNB.

In our calculations, we neglect the effects of disc outflows, which are expected to play an important role in hyperaccretion systems \citep[e.g.][]{Liu2008,Gu2015} and can significantly reduce the neutrino emission from NDAFs. Moreover, not all CCSNe are able to produce NDAFs, and the NDAF event rate depends on various progenitor properties. Therefore, the actual NDAF event rate is likely to be lower than that assumed in this work. As a result, the predicted DNNB flux and event rates presented in this work should be regarded as upper limits. Despite these uncertainties, our results demonstrate that the DNNB provides a promising probe of both NDAF physics and neutrino properties. Future neutrino detectors with improved sensitivity may be able to detect this background and place constraints on the nature of NDAFs, as well as on neutrino mass ordering.

\section*{Acknowledgements}

We thank Prof. Alexander Heger for providing us with pre-SN data and Prof. Shu Luo for very helpful comments and discussions. This work was supported by the National Key R\&D Program of China (Grant No. 2023YFA1607902), and the National Natural Science Foundation of China (Grant Nos. 12494572 and 12221003).

\section*{Data Availability}

The data underlying this article will be shared on reasonable request to the corresponding author.



\bibliographystyle{mnras}
\bibliography{ref} 








\bsp	
\label{lastpage}
\end{document}